\documentclass[sigconf]{acmart}
\AtBeginDocument{%
  }

\copyrightyear{2025}
\acmYear{2025}
\setcopyright{acmlicensed}
\acmConference[CIKM '25] {Proceedings of the 34th ACM International Conference on Information and Knowledge Management}{ November 10--14, 2025}{Seoul, Republic of Korea.}
\acmBooktitle{Proceedings of the 34th ACM International Conference on Information and Knowledge Management (CIKM '25), November 10--14, 2025, Seoul, Republic of Korea}
\acmDOI{10.1145/3746252.3761121}
\acmISBN{979-8-4007-2040-6/2025/11}

\usepackage{subcaption}
\usepackage{algorithm}
\usepackage{algorithmic}
\usepackage{balance}

\begin{document}

\title{Selective Mixup for Debiasing Question Selection in Computerized Adaptive Testing}

\author{Mi Tian}
\affiliation{%
  \institution{School of Computer Science and Information Engineering, Hefei University of Technology}
  \city{Hefei}
  \country{China}
}
\email{mitian@mail.hfut.edu.cn}

\author{Kun Zhang}
\affiliation{%
  \institution{School of Computer Science and Information Engineering, Hefei University of Technology}
  \city{Hefei}
  \country{China}
}
\email{zhang1028kun@gmail.com}
\authornote{Corresponding authors.}
\authornote{All authors are also with the Key Laboratory of Knowledge Engineering with Big Data, the Ministry of Education of China, Hefei, China.}

\author{Fei Liu}
\affiliation{%
  \institution{School of Computer Science and Information Engineering, Hefei University of Technology}
  \city{Hefei}
  \country{China}
}
\email{feiliu@mail.hfut.edu.cn}
\authornotemark[1]

\author{Jinglong Li}
\affiliation{%
  \institution{School of Computer Science and Information Engineering, Hefei University of Technology}
  \city{Hefei}
  \country{China}
}
\email{2023170684@mail.hfut.edu.cn}

\author{Yuxin Liao}
\affiliation{%
  \institution{School of Computer Science and Information Engineering, Hefei University of Technology}
  \city{Hefei}
  \country{China}
}
\email{yuxinliao314@gmail.com}

\author{Chenxi Bai}
\affiliation{%
  \institution{School of Computer Science and Information Engineering, Hefei University of Technology}
  \city{Hefei}
  \country{China}
}
\email{bcx@mail.hfut.edu.cn}

\author{Zhengtao Tan}
\affiliation{%
  \institution{School of Computer Science and Information Engineering, Hefei University of Technology}
  \city{Hefei}
  \country{China}
}
\email{1278985567@qq.com}

\author{Le Wu}
\affiliation{%
  \institution{School of Computer Science and Information Engineering, Hefei University of Technology}
  \city{Hefei}
  \country{China}
}
\email{lewu.ustc@gmail.com}

\author{Richang Hong}
\affiliation{%
  \institution{School of Computer Science and Information Engineering, Hefei University of Technology}
  \city{Hefei}
  \country{China}
}
\email{hongrc.hfut@gmail.com}

\renewcommand{\shortauthors}{Mi Tian et al.}

\begin{abstract}

Computerized Adaptive Testing is a widely used technology for evaluating examinees’ proficiency in online education platforms. By leveraging prior estimates of proficiency to select questions and updating the estimates iteratively based on responses, it enables personalized examinee modeling and has attracted substantial attention. Despite this progress, most existing works focus primarily on improving proficiency estimation accuracy, while overlooking the selection bias inherent in the adaptive process. Selection bias arises because the question selection is strongly influenced by the estimated proficiency, such as assigning easier questions to examinees with lower proficiency and harder ones to examinees with higher proficiency. Since the selection depends on prior estimation, this bias propagates into the diagnostic model, which is further amplified during iterative updates, leading to misaligned and biased predictions. Moreover, the imbalance in examinees’ historical interactions often exacerbates bias in diagnostic models. To address this issue, we propose a debiasing framework consisting of two key modules: Cross-Attribute Examinee Retrieval and Selective Mixup-based Regularization. First, we retrieve balanced examinees with relatively even distributions of correct and incorrect responses and use them as neutral references for biased examinees. Then, Mixup is applied between each biased examinee and its matched balanced counterpart under label consistency. This augmentation enriches the diversity of bias-conflicting samples and smooths selection boundaries. Finally, extensive experiments on two benchmark datasets with multiple advanced diagnosis models have been conducted. The results demonstrate that our method substantially improves the generalization ability of question selection.
\end{abstract}

\ccsdesc[500]{Applied computing~Computer-assisted instruction}

\keywords{Computerized Adaptive Testing, Bias Mitigation}


\maketitle

\section{Introduction}
 \begin{figure}[h]
  \centering
  \includegraphics[width=0.8\linewidth]{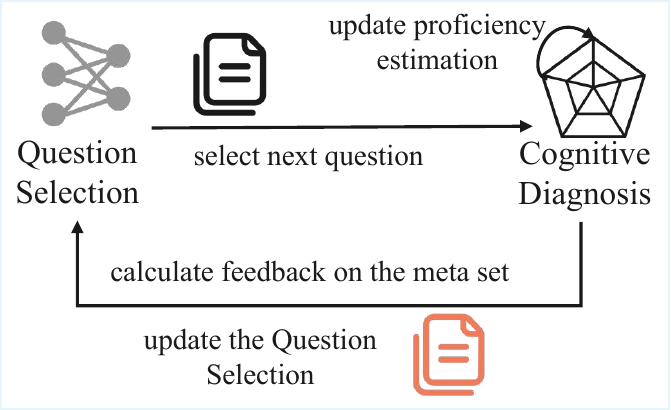}
  \caption{The training process of the data-driven Question Selection Module involves iterative updates guided by proficiency estimation and meta set feedback.} 
  \label{fig:intro} 
\end{figure}
Computerized Adaptive Testing (CAT) enables accurate proficiency estimation with fewer questions by iteratively selecting questions and updating examinee states~\cite{liu2024survey, choi2020development, yang2022adaptive}. While early CAT relied on heuristic strategies such as Maximum Fisher Information~\cite{lord2012mfi}, recent advances have shifted toward data-driven approaches that learn the Question Selection Module from large-scale interaction data~\cite{ghosh2021bobcat, wang2023gmocat, yu2024ucats}, achieving strong diagnostic performance. However, research on mitigating bias in CAT remains limited~\cite{kwon2023useraif, hort2024biassurvey}. Bias arising from data imbalance or adaptive selection may lead to the systematic assignment of easier or harder questions, thereby distorting proficiency estimates and compromising fairness. Prior work such as UserAIF~\cite{kwon2023useraif} alleviates bias by recalibrating question parameters, but it overlooks the effect of biased training data on the Question Selection Module itself.

The origins and consequences of biased selection remain insufficiently explored, leaving a gap in the development of fair and robust CAT systems. This work addresses this limitation by examining how biased selection arises during training and impacts both proficiency estimation accuracy and fairness, with a focus on mitigating bias within data-driven Question Selection Modules. Recent Question Selection Modules often adopt a bi-level optimization framework~\cite{ghosh2021bobcat, yu2024ucats}, where each examinee’s interactions are split into a support set and a meta set. The support set is used by the Cognitive Diagnosis Module (CDM) to estimate proficiency, while the meta set provides feedback for training the Question Selection Module. As illustrated in Figure~\ref{fig:intro}, at each step $t$, the Question Selection Module selects the next question based on the examinee’s history, the CDM updates proficiency after the response, and predictions on the meta set yield feedback to optimize the Question Selection Module. This process repeats for $T$ steps to iteratively refine question selection.

In this process, the meta set guides how the Question Selection Module is trained, aiming to align the distribution of selected questions with that of the meta set to improve proficiency estimation accuracy. In practice, however, the meta set is inherently imbalanced. As training progresses, the Question Selection Module overfits these skewed patterns, leading to divergence between the selected and the unbiased distributions, thereby exacerbating selection bias.

\begin{figure}[t]
  \centering
  \begin{subfigure}[b]{\linewidth}
      \centering
      \includegraphics[width=0.8\linewidth]{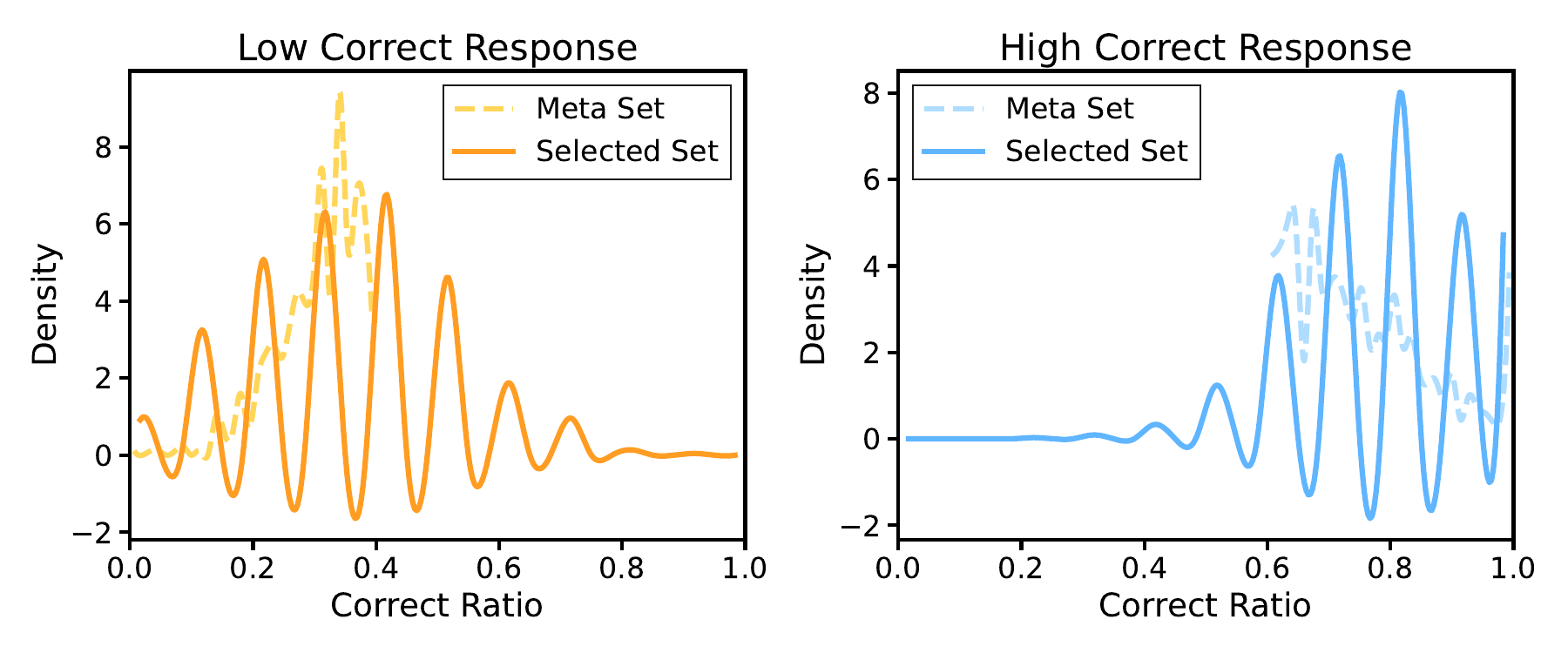}
      \caption{Distribution of the meta set and selected set.} 
      \label{fig:subfig_a_1} 
  \end{subfigure}
  \vspace{0.3cm} 
  \begin{subfigure}[b]{\linewidth}
      \centering
      \includegraphics[width=0.8\linewidth]{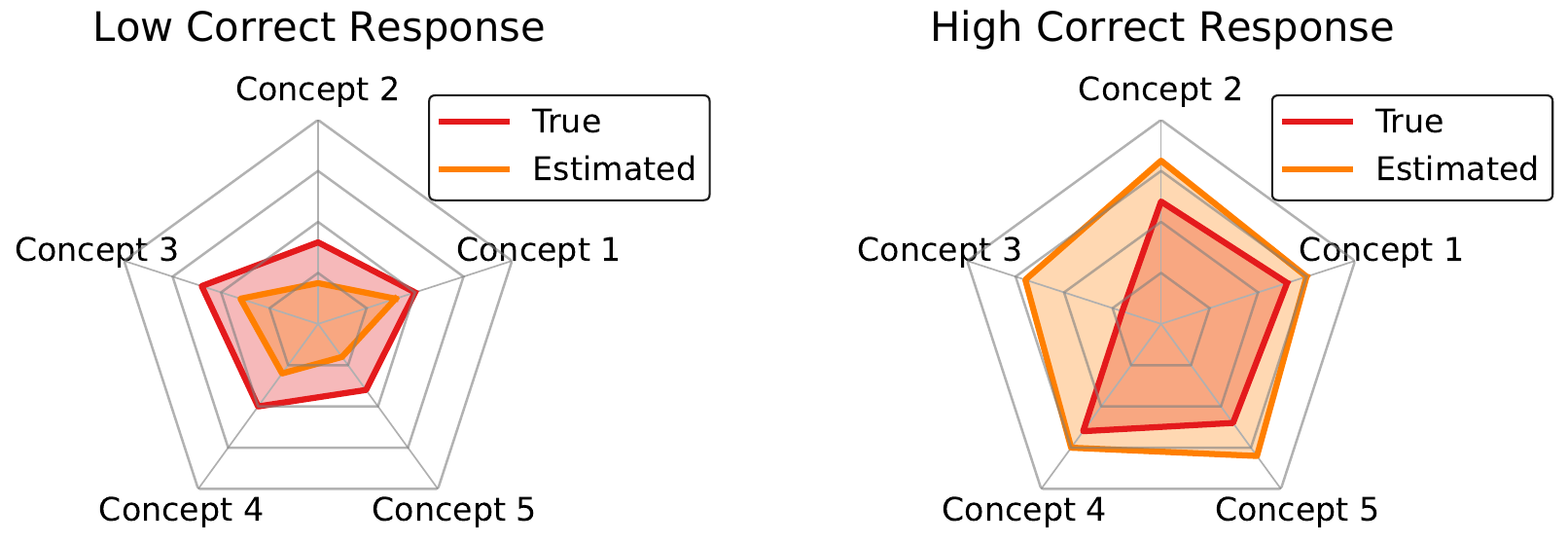}
      \caption{Biased proficiency estimation.} 
      \label{fig:subfig_b_1} 
  \end{subfigure}
  \caption{The imbalanced distribution of the meta set can induce shifts in the selection set, leading to biased estimates of examinee proficiency.}
  \label{fig:intro_2}
\end{figure}
Examinees differ in the proportion of their correct responses: some perform well, others poorly, while some are more balanced. We characterize each examinee by this correct-response ratio. As shown in Figure~\ref{fig:subfig_a_1}, the distribution of meta sets and selected data in NIPS-EDU dataset varies with this ratio\footnote{\url{https://eedi.com/projects/neurips-education-challenge}}, with density curves estimated by histograms and cubic spline smoothing. For high-ratio examinees, the meta set centers around 0.6 but shifts toward 1.0 after training, indicating a bias toward reinforcing strong performance. For low-ratio examinees, the distribution similarly shifts toward 0.0, reinforcing underperformance. Thus, an already imbalanced meta set becomes more skewed during training. This imbalance introduces \textbf{selection bias: the Question Selection Module tends to favor easier questions for high-ratio examinees and harder ones for low-ratio examinees, regardless of their true proficiency.} While data-driven CAT models achieve strong empirical performance, such biased behavior contradicts the core objective of CAT and leads to distorted outcomes. As illustrated in Figure~\ref{fig:subfig_b_1}, this results in unreliable estimates, systematically overestimating high performers while underestimating low performers.


Selection bias in data-driven CAT arises from imbalanced meta sets, where rare cases such as correct responses from low-ratio examinees or incorrect responses from high-ratio examinees are underrepresented.
This imbalance leads the Question Selection Module to overfit common patterns and weakens generalization. Since CAT jointly optimizes the CDM and the Question Selection Module, isolating the source of bias is challenging. To focus on selection bias, we fix the CDM and mitigate imbalance in the Question Selection Module by augmenting rare cases with interactions from balanced examinees sharing the same labels. Through selective Mixup of these samples, we obtain a more balanced and representative training distribution that guides the Question Selection Module toward smoother and less biased decision boundaries. Our contributions are summarized as follows:

\begin{itemize}
  \item Formally identify and analyze the problem of selection bias in data-driven CAT, tracing its origin to imbalanced meta-set distributions during training.  
  \item Propose a selective cross-attribute Mixup strategy that augments underrepresented interactions, enhances training diversity, and strengthens the generalization of the Question Selection Module.  
  \item Validate the proposed method through extensive experiments on two benchmark datasets, demonstrating its strong capability in mitigating bias. 
\end{itemize}

\section{Related Work}

\subsection{CAT and Bias in Cognitive Modeling}
Computerized Adaptive Testing (CAT) adaptively selects questions to estimate examinees’ knowledge proficiency~\cite{chang2015survey, liu2024survey}, typically consisting of a Cognitive Diagnosis Module (CDM) to infer proficiency and a Question Selection Module to determine the next question. Early CDMs were built on Item Response Theory~\cite{embretson2013irt} and its multidimensional extension~\cite{ackerman2014mirt}, while recent deep learning models such as NCDM~\cite{wang2020ncd}, KaNCD~\cite{wang2022kancd}, RCD~\cite{gao2021rcd}, and DCD~\cite{chen2023dcd} have greatly improved representation and proficiency estimation accuracy. For the Question Selection Module, heuristic approaches such as Maximum Fisher Information~\cite{lord2012mfi}, KL-based selection~\cite{chang1996kli}, MAAT~\cite{bi2020maat}, RAT~\cite{zhuang2022robust}, and BECAT~\cite{zhuang2023becat} are simple and interpretable, but limited in capturing complex interactions. More recent data-driven methods leverage bi-level optimization or reinforcement learning, including BOBCAT~\cite{ghosh2021bobcat}, NCAT~\cite{zhuang2022ncat}, GMOCAT~\cite{wang2023gmocat}, LACAT~\cite{cheng2024lacat}, and UATS~\cite{yu2024ucats}. However, data-driven methods often overlook bias in the data and selection process, which may compromise robustness in CAT.

Bias has been increasingly recognized in Cognitive Modeling, where predictions may differ systematically across sensitive attributes such as gender or socioeconomic status~\cite{xu2025fairwisa}. Methods like adversarial learning or causal reasoning~\cite{zhang2024faircd, zhang2024psrcf} have been proposed to mitigate such disparities. However, bias mitigation in CAT remains underexplored. In practice, CAT interaction data is inherently subject to selection bias, and directly training data-driven CAT models on such biased data can distort proficiency estimation and compromise system validity. Prior work such as UserAIF~\cite{kwon2023useraif} uses influence functions to identify less biased samples and recalibrate the question parameters in CDM.

\textbf{Our distinction.} Previous studies have focused on question-side bias, while our work targets the Question Selection Module, mitigating bias from adaptive selection and directly addressing it at its source rather than applying post-hoc debiasing, thereby improving fairness and robustness in CAT.

\subsection{Bias Mitigation Methods}
While machine learning systems can relieve humans from tedious tasks and perform complex calculations at high speed, they are only as reliable as the data on which they are trained~\cite{barocas2016big}. Since algorithms are not intentionally designed to incorporate bias, they often replicate or even amplify the spurious correlations present in real-world data, raising concerns regarding fairness and accountability~\cite{pedreshi2008discrimination, van2022overcoming}. To address these issues, a variety of strategies have been proposed. Invariant Risk Minimization aims to learn environment-invariant predictors~\cite{arjovsky2019irm}, while GroupDRO seeks to minimize worst-group risk but requires group labels~\cite{sagawa2019groupDRO}. Reweighting strategies adjust sample importance to counter imbalance~\cite{japkowicz2000reweight}, yet they only reshape optimization weights without enriching data diversity. In contrast, Mixup-based methods~\cite{zhang2018mixup} regularize models by interpolating samples and labels, with extensions such as Manifold Mixup~\cite{verma2019manifold} and Co-Mixup~\cite{kim2021co} improving hidden representation learning and pair selection. Building on this line of work, our method adopts a selective Mixup strategy to mitigate selection bias in CAT, leveraging its simplicity and effectiveness to enhance generalization.

\section{Preliminaries}
\subsection{Computerized Adaptive Testing}
Computerized Adaptive Testing is composed of two main components: a Cognitive Diagnosis Module for proficiency estimation and a Question Selection Module. In the following, we provide a detailed description of the structure of these two components.

\subsubsection{Cognitive Diagnosis Module}
The Cognitive Diagnosis Module (CDM) estimates an examinee’s proficiency $\boldsymbol{\theta}$ by fitting their historical responses and predicting performance on unseen questions. Formally, the model outputs the probability $p \in [0,1]$ that an examinee answers a given item correctly. A representative neural-network-based CDM is NCDM~\cite{wang2020ncd}, which leverages neural networks to capture complex interactions between examinees and items. The model incorporates several components: an examinee proficiency vector $\boldsymbol{\theta} \in (0,1)^{K}$, a question concept vector $\boldsymbol{Q}_{i} \in \{0,1\}^{1 \times K}$ indicating the required knowledge concepts, a knowledge difficulty vector $\boldsymbol{h}^{diff} \in (0,1)^{K}$, a scalar discrimination parameter $h^{disc} \in (0,1)$, and trainable neural network parameters.

\paragraph{Interaction Function.}
The first interaction layer is inspired by the Multidimensional Item Response Theory (MIRT)~\cite{ackerman2014mirt} model, and is defined as:
\begin{equation}\label{eq:eq1}
\boldsymbol{x} = \boldsymbol{Q}_{i} \circ (\boldsymbol{\theta} - \boldsymbol{h}^{diff}) \times h^{disc},
\end{equation}
where $\circ$ denotes the element-wise product.

The intermediate layers consist of two fully connected layers followed by an output layer:
\begin{subequations}\label{eq:eq2}
\begin{align}
\boldsymbol{f}_1 &= \varphi (\boldsymbol{W}_1 \boldsymbol{x}^T + \boldsymbol{b}_1), \label{eq:eq2a} \\
\boldsymbol{f}_2 &= \varphi (\boldsymbol{W}_2 \boldsymbol{f}_1 + \boldsymbol{b}_2), \label{eq:eq2b} \\
p &= \varphi (\boldsymbol{W}_3 \boldsymbol{f}_2 + \boldsymbol{b}_3), \label{eq:eq2c}
\end{align}
\end{subequations}

where $\varphi$ denotes the activation function, typically chosen as the sigmoid. The monotonicity assumption requires that the probability of correctly answering an item increases monotonically with respect to each dimension of an examinee’s knowledge proficiency. To enforce this property, NCDM adopts a simple yet effective strategy by constraining all elements of $\boldsymbol{W}_1$, $\boldsymbol{W}_2$, and $\boldsymbol{W}_3$ to be positive.



\paragraph{Loss Function.}
The cross-entropy loss is employed between the predicted probability $p \in [0,1]$ and the ground-truth response $y$:
\begin{equation}\label{eq:eq5}
\ell(y, p) = -\big( y \log p + (1-y) \log (1-p) \big),
\end{equation}
For simplicity, we denote the model output as $M(q;\boldsymbol{\theta}, \zeta_{q}, \zeta_{n}) = p$. 
Here, $q$ denotes a question, $\boldsymbol{\theta}$ represents the examinee’s proficiency, $\zeta_{q}$ corresponds to the question parameters (e.g., concept vector, difficulty, discrimination), 
and $\zeta_{n}$ denotes the trainable neural network parameters. 
After training, the estimated proficiency vector $\boldsymbol{\theta}$ serves as the diagnostic result, 
characterizing the examinee’s proficiency across the $K$ knowledge concepts.

\subsubsection{Question Selection Module}
Existing data-driven Question Selection Modules typically employ a bi-level optimization framework~\cite{ghosh2021bobcat}, where a selection network is trained on large-scale examinee data. The dataset $\mathcal{D}$ is partitioned into a support set $\mathcal{D}_u$ and a meta set $\mathcal{D}_m$, which are used to compute the loss or reward. As an illustrative example, we briefly introduce BOBCAT~\cite{ghosh2021bobcat}. At each step $t$, the selection network $\pi(\cdot)$ chooses a question $q_t \in \mathcal{D}_u$ based on the examinee’s current state $\boldsymbol{s}_t$, which is encoded from their interaction history $r = \{(q_1, y_1), \ldots, (q_{t-1}, y_{t-1})\}$, where $y_i \in \{0,1\}$ indicates whether the response was correct or incorrect.

The state $\boldsymbol{s}_t$ is represented as a vector with length equal to the number of questions, where each entry takes values in $\{-1, 0, 1\}$, corresponding to incorrect, unattempted, and correct responses, respectively. Formally, the selection process is defined as:
\begin{equation}
\label{eq:select}
q_t = \pi(\boldsymbol{s}_t, \mathcal{D}_u; \zeta_s),
\end{equation}
where $\zeta_s$ denotes the parameters of the selection network.

To train the selection network, we adopt a bi-level optimization framework. The outer loop updates only the selection network $\zeta_s$, since the Question Selection Module is assumed to be the primary source of bias. The CDM is kept fixed, with parameters $(\zeta_q, \zeta_n)$ pre-trained and frozen, while the proficiency representation $\boldsymbol{\theta}$ is updated in the inner optimization. After training, the selection network generates the next question for each examinee based solely on their current response state. The overall bi-level optimization procedure for data-driven question selection can be summarized as follows, outlining the inner and outer optimization.

\paragraph{Inner Optimization.}
Given a selected question sequence, the examinee’s proficiency $\boldsymbol{\theta}_i$ is estimated by minimizing the prediction loss on their responses in the support set $\mathcal{D}_u$, with fixed question parameters $(\zeta_q, \zeta_n)$. Through this process, $\boldsymbol{\theta}_i$ is updated to obtain the estimated proficiency vector $\boldsymbol{\theta}_i^*$:
\begin{equation}\label{eq6}
\boldsymbol{\theta}_i^* = \arg\min_{\boldsymbol{\theta}_i} 
\sum_{t=1}^{T} \ell\!\left(y_{i,q_t}, M\left(q_t;\boldsymbol{\theta}_i, \zeta_q, \zeta_n\right)\right),
\end{equation}
where $T$ denotes the number of selected questions.

\paragraph{Outer Optimization.}
The selection network $\zeta_s$ is updated by minimizing the prediction loss on the meta set $\mathcal{D}_m$, using the fixed proficiency $\boldsymbol{\theta}_i^*$ obtained from the inner optimization and the pre-trained CDM parameters $(\zeta_q, \zeta_n)$:
\begin{equation}\label{eq7}
\min_{\zeta_s} \frac{1}{N} \sum_{i=1}^N \sum_{q_j \in \mathcal{D}_m} 
\ell\!\left(y_{i,q_j}, M\left(q_j;\boldsymbol{\theta}_i^*, \zeta_q, \zeta_n\right)\right),
\end{equation}
where $N$ denotes the number of examinees. In our framework, the key contribution lies in introducing Mixup-based data augmentation within this outer optimization stage, thereby enhancing the robustness of the selection network and mitigating bias.

\begin{figure*}[ht]
  \centering
  \includegraphics[width=\textwidth]{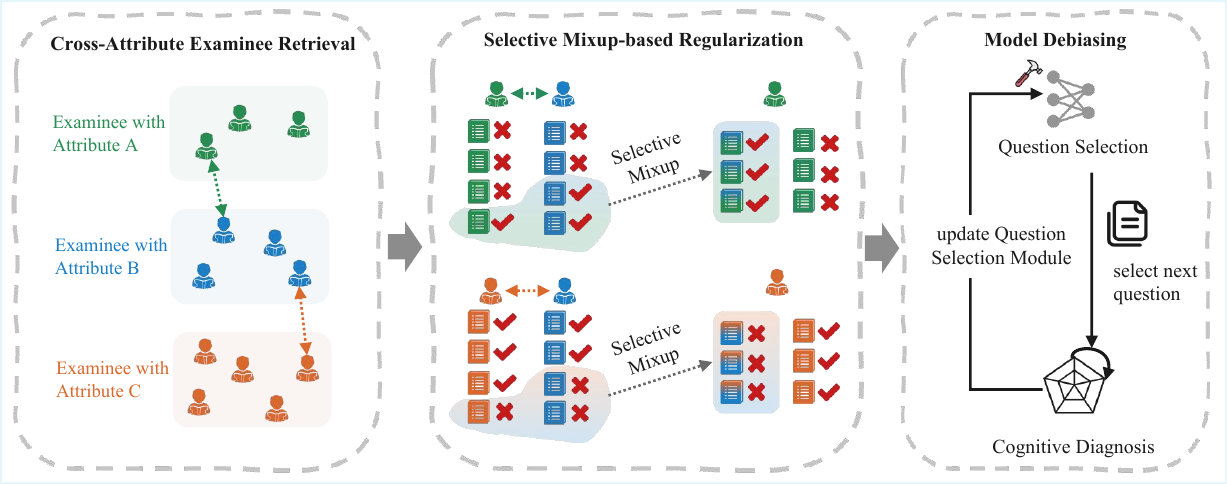}
  \caption{An overview of our debiasing framework for Computerized Adaptive Testing. The method consists of two stages: (1) \textbf{Cross-Attribute Examinee Retrieval}, where midperformers from $\mathcal{D}_m^{\text{B}}$ are retrieved as neutral references for examinees in $\mathcal{D}_m^{\text{A}}$ and $\mathcal{D}_m^{\text{C}}$ based on proficiency similarity; and (2) \textbf{Selective Mixup Regularization}, where Mixup is applied between retrieved midperformers and their matched counterparts (with identical labels) to generate synthetic training samples. This augmentation reduces reliance on bias-aligned patterns, thereby mitigating bias in the Question Selection Module.}
  \label{fig:method}
\end{figure*}
\subsection{Notation and Problem Setting}
We divide the examinees in $\mathcal{D}_m$ into three attributes, $\mathcal{D}_m^{\text{A}},\ \mathcal{D}_m^{\text{B}},\ \mathcal{D}_m^{\text{C}}$, according to their proportion of correct responses. These groups reflect different distributions in the meta set. Let $\mathcal{A} = \{\text{A}, \text{B}, \text{C}\}$ denote the set of attribute categories.
\begin{itemize}
   \item \textbf{A}: the proportion of correct responses lies in $[0, 0.4]$, where incorrect responses dominate.  
   \item \textbf{B}: the proportion of correct responses lies in $(0.4, 0.6]$, where correct and incorrect responses are approximately balanced.  
   \item \textbf{C}: the proportion of correct responses lies in $(0.6, 1.0]$, where correct responses dominate.  
\end{itemize}

We also define the label set as $\mathcal{Y} = \{0, 1\}$, where $0$ denotes an incorrect response and $1$ denotes a correct response. A group is defined as a unique combination of an attribute $a \in \mathcal{A}$ and a label $y \in \mathcal{Y}$. Based on this definition, interactions can be categorized into three types:
\begin{itemize}
  \item \textbf{Bias-aligned samples}: interactions that follow the dominant attribute pattern and are typically easier to learn, such as $\mathcal{D}_m^{\text{A}}$ with label $0$ or $\mathcal{D}_m^{\text{C}}$ with label $1$.  
  \item \textbf{Unbiased samples}: interactions from $\mathcal{D}_m^{\text{B}}$, where responses are relatively balanced and do not exhibit strong bias. These are considered neutral.  
  \item \textbf{Bias-conflicting samples}: interactions that diverge from the dominant attribute pattern and are harder to learn, for example, $\mathcal{D}_m^{\text{A}}$ with label~$1$ or $\mathcal{D}_m^{\text{C}}$ with label~$0$.
\end{itemize}

In CAT, an examinee’s true proficiency $\theta_i^*$ is assumed to remain constant, and traditional methods aim to make the final estimate $\theta_i^T$ closely approximate it. In contrast, our work focuses on mitigating selection bias in the data-driven Question Selection Module, producing more accurate and unbiased proficiency estimates while maintaining robustness under distributional shifts.

\section{Method}
This section introduces our approach to mitigating bias in data-driven CAT. The method targets selection bias in the meta set through a selective Mixup strategy guided by examinee attributes. By generating augmented samples from balanced response profiles, it enriches underrepresented cases and enables the Question Selection Module to learn smoother and less biased decision boundaries. The approach is simple yet effective and can be seamlessly integrated into data-driven question selection in CAT. In addition, it requires only minor modifications to existing training pipelines, making it computationally efficient and easy to implement. These properties ensure that the proposed method can be readily adopted in practical educational assessment systems.

\subsection{Overview} 
Our method mitigates selection bias in CAT through a two-stage framework: \textbf{Cross-Attribute Examinee Retrieval} and \textbf{Selective Mixup Regularization}, as illustrated in Figure~\ref{fig:method}. In the first stage, for each examinee in $\mathcal{D}_m^{\text{A}}$ or $\mathcal{D}_m^{\text{C}}$, the most similar counterpart is retrieved from $\mathcal{D}_m^{\text{B}}$ based on response patterns. In the second stage, the focus is on bias-conflicting samples, such as correct responses from $\mathcal{D}_m^{\text{A}}$ or incorrect responses from $\mathcal{D}_m^{\text{C}}$. To enrich these underrepresented cases, each examinee from $\mathcal{D}_m^{\text{B}}$ is paired with a counterpart from $\mathcal{D}_m^{\text{A}}$ or $\mathcal{D}_m^{\text{C}}$, and Mixup is applied only when their interaction labels are identical. The final training objective combines empirical loss on original data with regularization from synthetic samples, thereby enhancing the robustness and generalization of the Question Selection Module under biased meta distributions. In the following sections, we focus primarily on the outer optimization, where these debiasing strategies are applied.

\subsection{Cross-Attribute Examinee Retrieval}

In real-world CAT scenarios, training data often suffers from selection bias: examinees in $\mathcal{D}_m^{\text{A}}$ are more likely to respond incorrectly, while those in $\mathcal{D}_m^{\text{C}}$ tend to respond correctly. This leads to a dominance of bias-aligned interactions, which are easy to fit but prone to shortcut learning and reduced robustness. In contrast, bias-conflicting cases (e.g., correct responses from $\mathcal{D}_m^{\text{A}}$ or incorrect responses from $\mathcal{D}_m^{\text{C}}$) are rare yet crucial for achieving fairness and generalization. To mitigate this imbalance, we leverage examinees from $\mathcal{D}_m^{\text{B}}$, whose balanced response patterns serve as reliable references for augmenting underrepresented cases, thereby motivating our selective Mixup design.
Concretely, for each examinee $e_a$ with attribute $a \in \{\text{A}, \text{C}\}$, we retrieve the most similar examinee $e_b$ with attribute $b \in \text{B}$ based on their cognitive proficiency vectors $\boldsymbol{\theta}^*$, estimated by the CDM. The similarity between examinees is measured using the standard $L_2$ norm distance:
\vspace{-0.5em} 
\begin{equation}\label{eq:eq8}
\text{similarity}_{i,j} = \left\| \boldsymbol{\theta}_i^{*} - \boldsymbol{\theta}_j^{*} \right\|_2
\end{equation}
\vspace{-0.2em} 
where $\boldsymbol{\theta}_i^{*}$ and $\boldsymbol{\theta}_j^{*}$ denote the latent knowledge proficiency vectors of examinees $i$ and $j$, respectively.

This retrieval process serves two purposes: first, it anchors bias-conflicting interactions with semantically similar but statistically more balanced samples; and second, it provides the foundation for subsequent augmentation. By aligning examinees in $\mathcal{D}_m^{\text{A}}$ and $\mathcal{D}_m^{\text{C}}$ with appropriate references from $\mathcal{D}_m^{\text{B}}$, the method effectively bridges biased extremes and facilitates smoother decision boundary learning in the next stage.

\subsection{Selective Mixup-based Regularization}
To further mitigate selection bias, we propose a selective Mixup augmentation strategy~\cite{zhang2018mixup}. Unlike conventional training that relies solely on naturally occurring samples, Mixup generates synthetic interactions by interpolating instances, thereby substantially enriching the diversity of bias-conflicting data. This is particularly important in CAT, where the scarcity of such cases often causes the Question Selection Module to overfit bias-aligned patterns and neglect minority interactions. In this context, Mixup offers several advantages: it smooths decision boundaries, reduces overfitting, and improves generalization when test distributions differ from training distributions. Moreover, by diluting the influence of spurious correlations in the original data, Mixup encourages the model to learn more balanced and robust representations. Through this augmentation, the Question Selection Module is guided toward fairer and more reliable predictions, enhancing robustness and generalization under distribution shifts.
\paragraph{Bias-Conflicting Sample Generation} 
Given $q_i \in \{\mathcal{D}_m^{\text{A}}, \mathcal{D}_m^{\text{C}}\}$, its most similar counterpart in terms of proficiency is first retrieved from $\mathcal{D}_m^{\text{B}}$ in the previous stage. From this matched examinee in $\mathcal{D}_m^{\text{B}}$, an interaction $q_j$ with the same true label $y$ is sampled and paired with $q_i$, upon which synthetic bias-conflicting samples are generated through convex combination. With such paired questions $(q_i, q_j)$, we interpolate their parameters to generate diverse synthetic questions and enrich the training distribution, forming the synthetic set $\mathcal{D}_{\text{syn}}$. The resulting representation Mixup is defined as weighted interpolation of key parameters, where $\lambda \sim \text{Beta}(\alpha, \alpha)$ with $\alpha \in [0,1]$ controlling intensity, as shown below:
\begin{subequations}\label{eq:mixup}
\begin{align}
\boldsymbol{Q}_{i,j} &= \lambda \boldsymbol{Q}_{i} + (1-\lambda)\boldsymbol{Q}_{j}, 
\label{eq:mixup_a} \\
\boldsymbol{h}_{i,j}^{\text{diff}} &= \lambda \boldsymbol{h}^{\text{diff}}_{i} + (1-\lambda)\boldsymbol{h}_{j}^{\text{diff}}, 
\label{eq:mixup_b} \\
h_{i,j}^{\text{disc}} &= \lambda h_{i}^{\text{disc}} + (1-\lambda)h_{j}^{\text{disc}},
\label{eq:mixup_c}
\end{align}
\end{subequations}



\paragraph{Synthetic Feature Construction}
The synthesized features are fed into the CDM, which computes the corresponding response probability as $M(q_{i,j}; \boldsymbol{\theta}_i^*, \zeta_{q_{i,j}}, \zeta_n)$. Since Mixup is applied only to samples with the same label, the synthetic sample keeps that label. The loss for these augmented samples is defined as:
\begin{equation}
\label{eq:l_fake}
\mathcal{L}_{\text{syn}} = 
\frac{1}{N} \sum_{i=1}^N  \sum_{q_{i,j} \in \mathcal{D}_{syn}} 
\ell\!\left(y_{i,q_{i,j}}, M\!\left(q_{i,j}; \boldsymbol{\theta}_i^*, \zeta_{q_{i,j}}, \zeta_n\right)\right),
\end{equation}

\paragraph{Debiasing Optimization}
The final outer-level optimization objective, extending Eq.~\eqref{eq7}, is expressed as a combination of empirical risk minimization and synthetic sample regularization:
\begin{gather}
\mathcal{L}_{\text{emp}} = 
\frac{1}{N} \sum_{i=1}^N \sum_{q_{j} \in \mathcal{D}_m} 
\ell\!\left(y_{i,q_{j}}, M\!\left(q_{j}; \boldsymbol{\theta}_i^*, \zeta_{q_{j}}, \zeta_n\right)\right), \label{eq:l_emp}\\[6pt]
\min_{\zeta_s} \mathcal{L}_{\text{final}} = \mathcal{L}_{\text{emp}} + \omega \mathcal{L}_{\text{syn}},
\label{eq:final_objective}
\end{gather}
where $\omega$ is a regularization coefficient that balances authentic and synthetic samples.

The dual-objective design encourages the model to preserve strong predictive performance on authentic data while enhancing robustness to bias-conflicting cases. In the outer-level optimization, the Question Selection Module parameters $\zeta_s$ are updated via gradient descent on the combined loss. A complete overview of the training process, incorporating Mixup-based augmentation and bias-aware optimization, is provided in Algorithm~\ref{alg:train}, which summarizes the key steps and illustrates how these components are integrated into the overall framework.
\begin{algorithm}[!h]
    \caption{Training Process of Debiased CAT}
    \label{alg:train}
    \begin{algorithmic}[1]
        \STATE Initialize examinee proficiency parameters $\boldsymbol{\theta}$ in the Cognitive Diagnosis Module with pre-trained values $(\zeta_{q}, \zeta_{n})$, and initialize $\zeta_s$ for the Question Selection Module. Specify learning rates $\alpha$ and $\beta$, the number of gradient descent steps $K$ for inner-level optimization, and the number of test steps $T$.
        \WHILE{not converged}
            \STATE Randomly sample a mini-batch of examinees $\mathcal{D}$ with support and meta sets $\{\mathcal{D}_u, \mathcal{D}_m\}$, where $\mathcal{D}_m = \mathcal{D}_m^{\text{A}} \cup \mathcal{D}_m^{\text{B}} \cup \mathcal{D}_m^{\text{C}}$.
            \FOR{$t = 1$ to $T$}
                \STATE Represent $s_{i,t}$ using the examinee's response history.
                \STATE Select question $q_{i,t}$ for each examinee based on current state $s_{i,t}$ (Eq.~\eqref{eq:select}).
                \STATE \textbf{Inner-level update:} update $\boldsymbol{\theta}_i$ by gradient descent with learning rate $\alpha$ on the prediction loss in Eq.~\eqref{eq6}, repeated for $K$ steps, to obtain $\boldsymbol{\theta}_i^*$.
                \STATE For examinees in $\mathcal{D}_m^{\text{A}}$ and $\mathcal{D}_m^{\text{C}}$, retrieve the most similar examinees from $\mathcal{D}_m^{\text{B}}$ using Eq.~\eqref{eq:eq8}.
                \STATE Apply selective Mixup regularization using Eqs.~\eqref{eq:mixup} and \eqref{eq:l_fake}, generating synthetic bias-conflicting samples that enrich the training distribution.
                \STATE \textbf{Outer-level update:} $\zeta_s \gets \zeta_s - \beta \nabla_{\zeta_s} \mathcal{L}_{\text{final}}$ (Eq.~\eqref{eq:final_objective}).
            \ENDFOR
        \ENDWHILE
        \RETURN Estimated proficiencies $\{\boldsymbol{\theta}_i^*\}_{i=1}^N$ and debiased Question Selection Module $\pi$.
    \end{algorithmic}
\end{algorithm}

\paragraph{Rationale and Impact}
By interpolating between samples from attribute A or C and their corresponding references from attribute B, our approach introduces balanced, label-consistent synthetic data into the training process. This prevents the Question Selection Module from overfitting to dominant response patterns and encourages it to capture more nuanced relationships between question difficulty and proficiency. As a result, the model becomes less reliant on biased historical distributions, thereby improving robustness and mitigating bias across diverse examinees.

\section{Experiments}
In this section, we first introduce the dataset and experimental setup. Then, we conduct extensive experiments on real-world benchmark datasets. The code used in our experiments is publicly available at: \href{https://github.com/xiaoluobouu/MDCAT}{\underline{https://github.com/xiaoluobouu/MDCAT}}.

\subsection{Experimental Settings}
In this section, we present the datasets, selected baselines, and the application of CAT. All models are implemented in PyTorch and trained on an NVIDIA RTX 4090 GPU to ensure efficient and consistent experimentation.

\textbf{Dataset Description.}
We conduct experiments on two widely used educational datasets. The ASSIST0910 dataset\footnote{\url{https://sites.google.com/site/assistmentsdata/home/assistment2009-2010-data}} contains examinee practice logs from the ASSISTments online tutoring system, focusing on mathematics and knowledge concept problems. The NIPS-EDU dataset\footnote{\url{https://eedi.com/projects/neurips-education-challenge}}, released in the NeurIPS 2020 Education Challenge, consists of examinee responses collected from the Eedi learning platform, where we use the interaction data from Task 3 and Task 4. For ASSIST0910, we remove records with missing values and exclude examinees with fewer than 40 interactions. The dataset statistics, including the number of examinees, questions, concepts, and interactions, are summarized in Table~\ref{tab:dataset_statistics}.
\begin{table}[t]
  \caption{The statistics of the datasets.}
  \label{tab:dataset_statistics}
  \centering
  \begin{tabular}{lcc}
    \toprule
    Dataset        & ASSIST0910     & NIPS-EDU \\
    \midrule
    Examinees      &1, 360       & 4,918   \\
    Questions          &17,372         & 948      \\
    Concepts       &119         & 86      \\
    Interactions   &241,156     & 1,382,727  \\
    \bottomrule
  \end{tabular}
\end{table}

\textbf{Partition.} We randomly split examinees into training (60\%), validation (20\%), and testing (20\%) sets. For evaluation, we perform 5-fold validation on each dataset, with each round using a different random seed. Within each round, the questions answered by each examinee are further divided into a support set ($\mathcal{D}_u$, 80\%) and a meta set ($\mathcal{D}_m$, 20\%), and to mitigate overfitting, these partitions are regenerated at the beginning of each training epoch. In the independent and identically distributed (IID) setting, the test split is also sampled randomly. 
In contrast, the out-of-distribution (OOD) setting follows a different partition strategy, which is described in detail in Section~\ref{sec:performance_in_ood_settings}.

\begin{table*}
  \caption{Performance of different CAT models under IRT- and NCDM-based cognitive diagnosis on the unbiased ASSIST0910 and NIPS-EDU datasets. Results are reported for Metrics@5 (top block) and Metrics@10 (bottom block). \emph{Worst} refers to worst-group accuracy, and \emph{Avg.} refers to overall average accuracy.}
  \label{tab:ood_per}
  \centering
  \resizebox{\textwidth}{!}{
  \begin{tabular}{lcccccccc|cccccccc}
    \toprule
    \multicolumn{1}{c}{Dataset} & \multicolumn{8}{c}{ASSIST0910} & \multicolumn{8}{c}{NIPS-EDU} \\
    \cmidrule(lr){2-9} \cmidrule(lr){10-17}
    CDM-CAT & \multicolumn{2}{c}{IRT-BOBCAT} & \multicolumn{2}{c}{IRT-UATS} & \multicolumn{2}{c}{NCDM-BOBCAT} & \multicolumn{2}{c}{NCDM-UATS} 
            & \multicolumn{2}{c}{IRT-BOBCAT} & \multicolumn{2}{c}{IRT-UATS} & \multicolumn{2}{c}{NCDM-BOBCAT} & \multicolumn{2}{c}{NCDM-UATS} \\
    \cmidrule(r){2-3} \cmidrule(r){4-5} \cmidrule(r){6-7} \cmidrule(r){8-9} 
    \cmidrule(r){10-11} \cmidrule(r){12-13} \cmidrule(r){14-15} \cmidrule(r){16-17}
    Metrics & Worst & Avg. & Worst & Avg. & Worst & Avg. & Worst & Avg. & Worst & Avg. & Worst & Avg. & Worst & Avg. & Worst & Avg. \\
    \midrule
    \multicolumn{17}{c}{\textbf{Metrics@5}} \\
    \midrule
    ERM      & 0.3824 & 0.6118 & 0.3400 & 0.6047 & 0.3662 & 0.6007 & 0.3268 & 0.5992 &
               0.4068 & 0.6383 & 0.5064 & 0.6353 & 0.3618 & 0.6339 & \textbf{0.4479} & 0.6439 \\
    IRM      & 0.3489 & 0.6045 & 0.3280 & 0.5964 & 0.2983 & 0.5883 & 0.2860 & 0.5813 &
               0.4140 & 0.6376 & 0.4450 & \textbf{0.6429} & 0.2916 & 0.6268 & 0.4098 & 0.6456 \\
    GroupDRO & 0.5077 & 0.6303 & 0.4057 & 0.6134 & 0.3279 & 0.5949 & 0.3066 & 0.5894 &
               0.3968 & \textbf{0.6384} & 0.4512 & 0.6427 & 0.2312 & 0.6241 & 0.4151 & 0.6432 \\
    Reweight & 0.5328 & \textbf{0.6395} & 0.3967 & 0.6142 & 0.4793 & \textbf{0.6152} & 0.4362 & 0.6107 &
               0.5699 & 0.6363 & 0.4832 & 0.6272 & 0.4040 & 0.6374 & 0.4172 & 0.6468 \\
    Ours     & \textbf{0.5470} & 0.6393 & \textbf{0.4382} & \textbf{0.6178} & \textbf{0.4989} & 0.6116 & \textbf{0.4823} & \textbf{0.6124} &
               \textbf{0.5701} & 0.6361 & \textbf{0.5185} & 0.6371 & \textbf{0.4250} & \textbf{0.6386} & 0.4315 & \textbf{0.6473} \\
    \midrule
    \multicolumn{17}{c}{\textbf{Metrics@10}} \\
    \midrule
    ERM  & 0.4088 & 0.6147 & 0.2885 & 0.5904 & 0.3206 & 0.5970 & 0.3188 & 0.5965 &
            0.3383 & 0.6191 & 0.4020 & 0.6254 & 0.2847 & 0.6263 & 0.3512 & 0.6399 \\
    IRM  & 0.3890 & 0.6136 & 0.2491 & 0.5847 & 0.3277 & 0.5994 & 0.3079 & 0.5920 &
            0.3245 & 0.6224 & 0.3532 & 0.6333 & 0.3056 & 0.6279 & \textbf{0.3704} & 0.6392 \\
    GroupDRO & 0.4370 & 0.6285 & 0.3424 & 0.6011 & 0.3600 & 0.6003 & 0.2796 & 0.5875 &
            0.2962 & 0.6253 & 0.3402 & \textbf{0.6340} & 0.2577 & 0.6247 & 0.3363 & 0.6424 \\
    Reweight & 0.3625 & 0.6299 & 0.3356 & 0.6050 & \textbf{0.4551} & 0.6102 & 0.4530 & 0.6091 &
            0.3133 & 0.6299 & 0.4496 & 0.6243 & 0.3446 & 0.6374 & 0.3458 & 0.6416 \\
    Ours & \textbf{0.4489} & \textbf{0.6320} & \textbf{0.3854} & \textbf{0.6095} & 0.4426 & \textbf{0.6117} & \textbf{0.4753} & \textbf{0.6155} &
            \textbf{0.4489} & \textbf{0.6312} & \textbf{0.4674} & 0.6268 & \textbf{0.3520} & \textbf{0.6381} & 0.3517 & \textbf{0.6436} \\
    \bottomrule
  \end{tabular}}
\end{table*}
\textbf{Evaluation.} We evaluate all methods on their ability to predict binary-valued responses in the meta set $\mathcal{D}_m$, using two complementary metrics: worst-group accuracy (\textbf{Worst}) and overall average accuracy (\textbf{Avg.}). The \textbf{Worst} metric reflects model robustness by focusing on the most disadvantaged groups, while the \textbf{Avg.} metric captures the overall prediction performance across all examinees. 

\textbf{Baseline Methods.} For CDMs, we adopt the 1PL IRT model~\cite{embretson2013irt} and NCDM~\cite{wang2020ncd}, both implemented in the EduStudio\footnote{\url{https://edustudio.ai/}} framework~\cite{wu2025edustudio}. For CAT frameworks, we consider data-driven approaches that formulate CAT as a bi-level optimization problem solved via meta-learning, including BOBCAT~\cite{ghosh2021bobcat} and UATS~\cite{yu2024ucats}. BOBCAT introduces a learnable Question Selection Module, whereas UATS employs a differentiable hierarchical optimization framework to improve efficiency and adaptability. To better analyze bias in question selection, we freeze the question parameters of the CDM during outer-loop optimization and initialize them with pre-trained values.






To validate the effectiveness of our method, we compare it against the following debiasing baselines, all implemented within the same bi-level training framework and adapted to the CAT setting to ensure consistent evaluation across different situations.

\textbf{ERM} (Empirical Risk Minimization) minimizes the average prediction loss over the entire meta set without explicitly addressing distributional bias. It serves as a standard baseline but often overfits to dominant bias-aligned patterns.

\textbf{IRM}~\cite{arjovsky2019irm} (Invariant Risk Minimization) encourages feature representations that yield consistent predictions across environments. In our setting, we simulate multiple environments by sampling examinees with different attribute values (A, B, and C). This approach ensures that the Question Selection Module performs robustly across these diverse distributions.

\textbf{GroupDRO}~\cite{sagawa2019groupDRO} (Group Distributionally Robust Optimization) improves worst-case performance across predefined groups. It optimizes the Question Selection Module by minimizing the maximum group-wise loss. This strategy helps prevent collapse on underrepresented or difficult subpopulations.

\textbf{Reweight}~\cite{japkowicz2000reweight} mitigates imbalance by assigning larger weights to bias-conflicting samples during training. This guides the model to better capture underrepresented patterns and reduces reliance on shortcut signals from bias-aligned data.


\textbf{Parameter Settings.}
In NCDM, the two hidden layers contain 128 and 64 neurons, which control the network’s capacity and complexity and allow it to capture nonlinear relationships between examinees and items. The $\alpha$ parameter of the Beta distribution is chosen from $\{0.2, 0.4, 0.6, 0.8, 1.0\}$, and the mixing coefficient $\omega$ in the Mixup loss is selected from the same range to ensure consistent regularization strength. To approximate each examinee’s proficiency, inner optimization updates the latent proficiency vector $K$ times, with $K=5$ in our experiments to balance accuracy and computational cost.

\subsection{Performance in OOD Settings}\label{sec:performance_in_ood_settings}
\textbf{Motivation.} 
In most existing data-driven CAT studies, the meta set used to evaluate question selection performance is directly sampled from each examinee’s original interaction log. This approach implicitly assumes that the observed response distribution faithfully reflects the examinee’s true exposure and proficiency profile. In practice, however, this assumption is problematic. Because CAT adaptively selects questions based on estimated proficiency, examinees are only exposed to a limited and biased subset of the entire question pool. As a result, the meta set often captures selection-induced artifacts rather than true proficiency, leading to biased and unrepresentative evaluation.

To evaluate the effectiveness of our debiasing strategy, we design an \textit{Out-of-Distribution (OOD)} test setting. The training and validation splits remain identical to the IID setup; only the test partition is modified to construct the OOD scenario. Specifically, our protocol enforces a 1:1 balance of correct and incorrect responses in the meta set for each examinee, with the remaining interactions assigned to the support set. Examinees who cannot satisfy this balancing constraint are excluded. This design preserves the standard bi-level framework while deliberately introducing a distribution shift. 
As a result, it provides a more rigorous and interpretable benchmark for assessing model robustness and generalization, 
especially when evaluation is conducted under biased conditions.

As shown in Table~\ref{tab:ood_per}, our method demonstrates clear advantages in mitigating selection bias across both datasets and diagnostic settings. When the number of selected questions increases from 5 to 10, the bias becomes more pronounced. The \emph{Worst} metric drops substantially while the \emph{Avg.} metric remains relatively stable, indicating that longer testing horizons disproportionately disadvantage underrepresented groups. The improvements of our approach are concentrated on the \emph{Worst} metric, with relative gains ranging from 2.7\% to 24.64\%, whereas the \emph{Avg.} metric increases by only about 0.1\%--1.05\%. For instance, on ASSIST0910 with NCDM-BOBCAT at Metrics@5, our method improves the \emph{Worst} from 0.3662 (ERM) to 0.4989; on ASSIST0910 with IRT-BOBCAT at Metrics@10, it raises the \emph{Worst} from 0.4088 (ERM) to 0.4489. For the \emph{Avg.} metric, the largest gain is observed on ASSIST0910 at Metrics@10, where our method improves from 0.6091 to 0.6155, achieving a relative improvement of 1.05\%. Overall, these results show that our method primarily enhances robustness by improving the performance of disadvantaged groups, while maintaining competitive overall accuracy. This aligns with the primary design goal of our debiasing framework, which is to reduce systematic disparities across different groups. 
At the same time, it ensures that overall performance is not sacrificed, maintaining the effectiveness of the model.

\subsection{Performance in IID Settings}

As shown in Table~\ref{tab:ood_per} and Table~\ref{tab:iid_per}, debiasing inevitably introduces a trade-off between IID and OOD performance. While methods such as Reweight and our approach effectively mitigate bias and improve OOD generalization, they cannot surpass ERM or GroupDRO under IID settings, which are naturally more favorable to models optimized without debiasing constraints. For example, IRM achieves the best average performance on ASSIST0910 (0.6939), whereas ERM attains the highest average on NIPS-EDU (0.6930). GroupDRO also yields relatively strong IID results across both datasets. By contrast, our method prioritizes robustness: it consistently outperforms the Reweight baseline in both \emph{Worst} and \emph{Avg.} accuracy under OOD conditions, while remaining competitive in-distribution. These results confirm that debiasing methods, including ours, trade a small amount of IID accuracy for substantially stronger OOD generalization. 
This finding clearly underscores the inherent balance and trade-offs involved in robustness-oriented learning.

\begin{table}[t]
\centering
\caption{IID results are included as a reference to illustrate performance under in-distribution conditions.}
\label{tab:iid_per}
\begin{tabular}{lllll}
\hline
Dataset    & \multicolumn{2}{c}{ASSIST0910} & \multicolumn{2}{c}{NIPS-EDU}   \\ 
CDM-CAT    & \multicolumn{2}{c}{IRT-BOBCAT} & \multicolumn{2}{c}{IRT-BOBCAT} \\
Metrics@10 & Worst          & Avg.          & Worst          & Avg.          \\
\hline
ERM        & 0.4609         & 0.6895 & 0.3937         & \textbf{0.6930} \\
IRM        & 0.4248         & \textbf{0.6939}         & 0.3854         & 0.6928          \\
GroupDRO   & 0.5104         & 0.6630          & 0.3535         & 0.6922          \\
Reweight   & 0.4179         & 0.6155          & 0.3576         & 0.6622          \\
Ours       & \textbf{0.5310} & 0.6599         & \textbf{0.4911} & 0.6695          \\ \hline
\end{tabular}
\end{table}

\subsection{Hyperparameter Analysis}
\begin{figure}[t]
    \centering
    \includegraphics[width=0.5\textwidth]{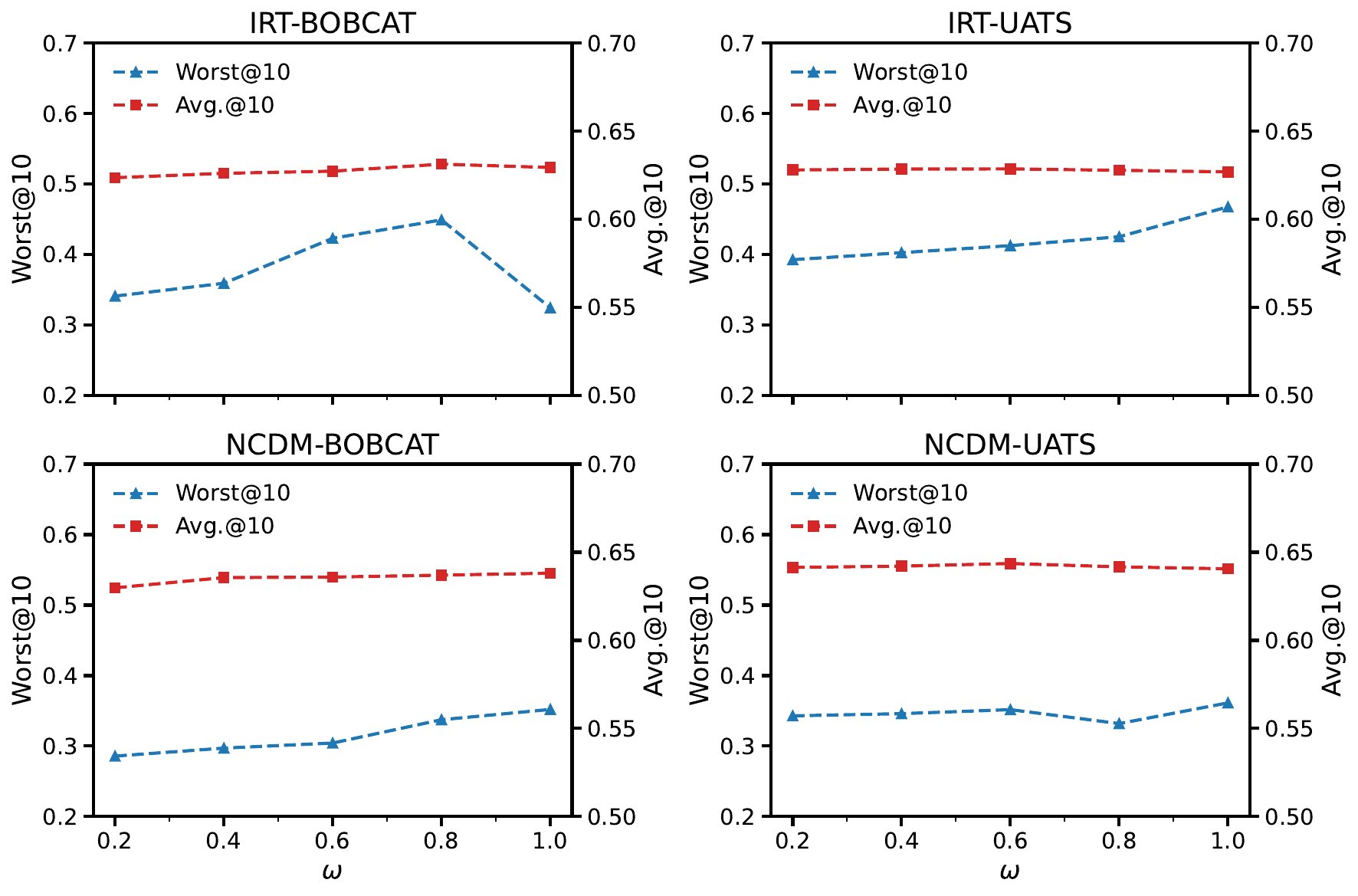} 
    \caption{Hyperparameter sensitivity of Mixup regularization weight $\omega$ on NIPS-EDU dataset.}
    \label{fig:hyp_analysis}
\end{figure}

We conduct a hyperparameter analysis of the Mixup regularization weight $\omega$ on NIPS-EDU dataset, examining its effect on both \emph{Worst} and \emph{Avg.} accuracy across two Cognitive Diagnosis Models and two CAT frameworks. As shown in Figure~\ref{fig:hyp_analysis}, different CDM–CAT combinations exhibit distinct behaviors: NCDM-BOBCAT and IRT-UATS display a steady upward trend as $\omega$ increases, indicating that stronger regularization consistently benefits these settings, whereas IRT-BOBCAT and NCDM-UATS follow a rise-then-fall pattern, with performance improving up to a moderate $\omega$ but deteriorating when regularization becomes excessive. This suggests that while Mixup is generally effective, overly strong interpolation may introduce noise that counteracts its debiasing effect. Overall, the curves remain relatively stable across a wide range of $\omega$ values, showing that our method is not highly sensitive to this hyperparameter. Notably, the \emph{Avg.} and \emph{Worst} metrics do not always peak at the same $\omega$, revealing a trade-off between fairness and generalization that necessitates careful calibration to achieve balanced improvements.

\subsection{Ablation Study}
We conduct ablation studies to evaluate three Mixup strategies: \textit{Mixup-Self}, \textit{Mixup-Inner}, and \textit{Mixup-B}. In \textit{Mixup-Self}, interactions are interpolated within the same examinee, introducing synthetic variation into an individual’s profile and occasionally improving robustness. \textit{Mixup-Inner} interpolates the most similar examinees within the same attribute, smoothing local representations and reducing overfitting to individual-specific patterns. In contrast, \textit{Mixup-B} interpolates biased examinees (e.g., from attribute A or C) with balanced counterparts (attribute B). By explicitly generating bias-conflicting interactions, this strategy enriches underrepresented patterns. It also provides a more effective and principled way to mitigate selection bias, thereby improving model robustness.

\begin{figure}[t]
  \centering
  \begin{subfigure}[b]{\linewidth}
      \centering
      \includegraphics[width=\linewidth]{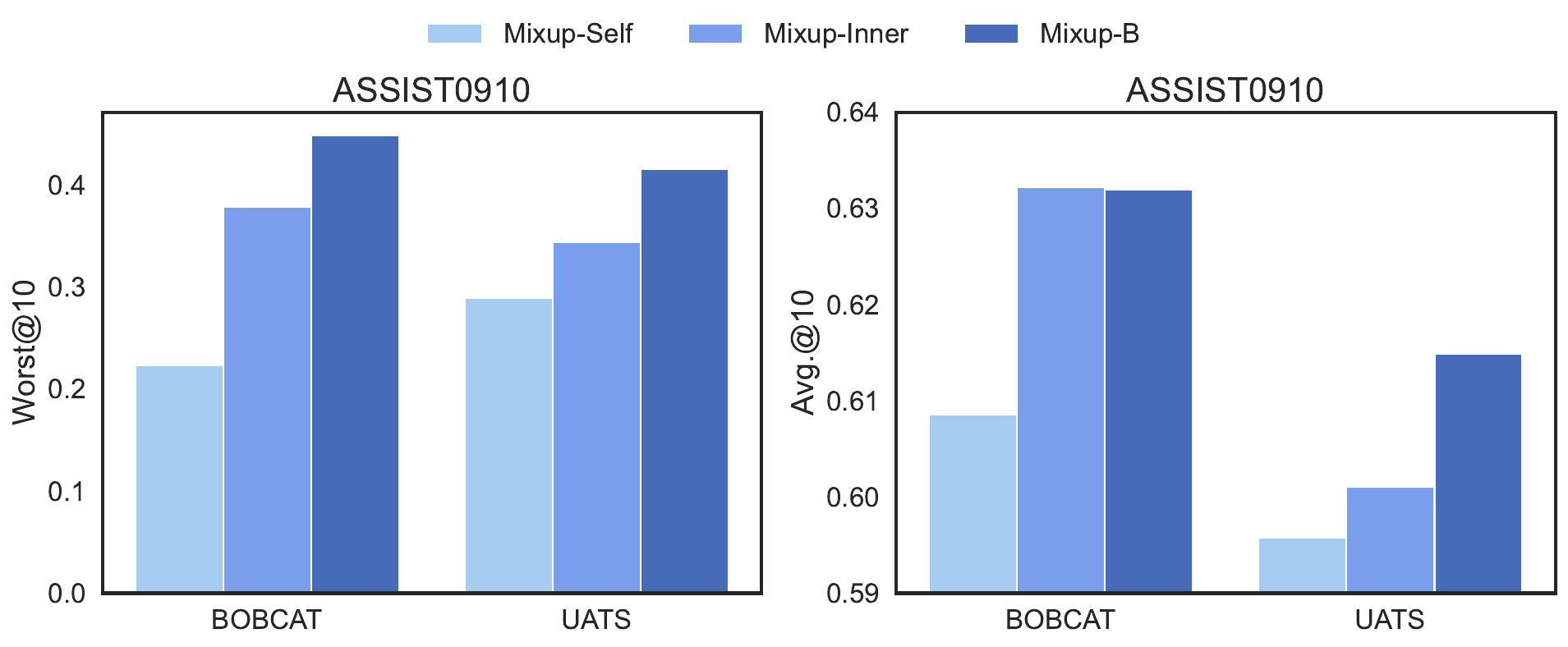}
      \caption{Ablation results of different Mixup strategies on ASSIST0910.} 
      \label{fig:subfig_a} 
  \end{subfigure}
  \vspace{0.3cm} 
  \begin{subfigure}[b]{\linewidth}
      \centering
      \includegraphics[width=\linewidth]{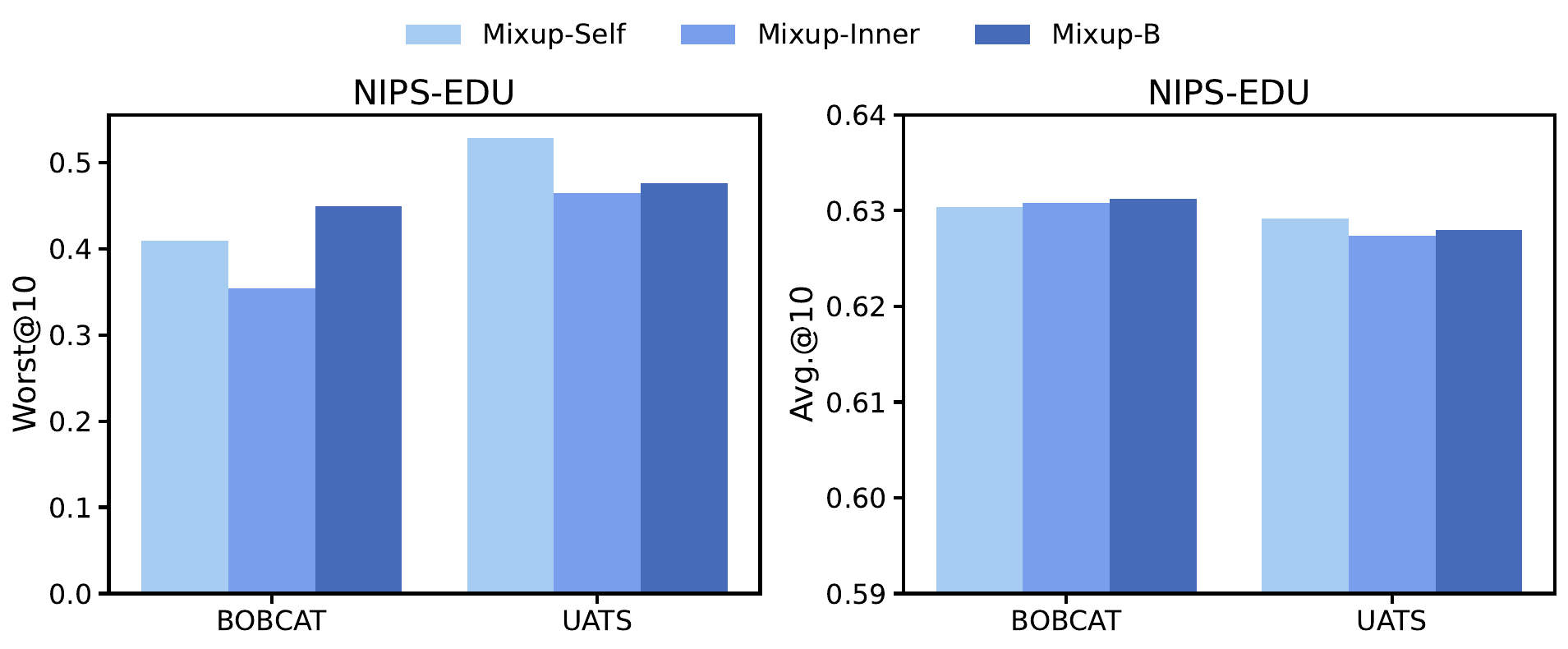}
      \caption{Ablation results of different Mixup strategies on NIPS-EDU.} 
      \label{fig:subfig_b} 
  \end{subfigure}
  
  \caption{Ablation results of different Mixup strategies (Mixup-Self, Mixup-Inner, and Mixup-B) on ASSIST0910 and NIPS-EDU in terms of Worst@10 and Avg.@10.}
  \label{fig:ablation}
\end{figure}

As shown in Figure~\ref{fig:ablation}, \textit{Mixup-B} consistently achieves the best performance, particularly on the large-scale ASSIST0910 dataset, where the larger question pool provides greater diversity and amplifies the benefits of cross-attribute mixing. On NIPS-EDU, although the dataset is smaller, the same trend persists, with \textit{Mixup-B} delivering the most stable improvements on both \emph{Worst} and \emph{Avg.}. These results demonstrate that the advantages of cross-attribute interpolation are not limited to a single dataset but generalize well across different CAT settings. At the same time, the comparison shows that \textit{Mixup-Self} and \textit{Mixup-Inner} can yield localized gains, but their instability makes them less reliable. Overall, the findings confirm that bias-conflicting augmentation is central to improving robustness and suggest promising directions for hybrid designs that combine multiple Mixup strategies in future work.

\subsection{Mitigating Distribution Shift in Question Selection}

\begin{figure}[t]
    \centering
    \includegraphics[width=0.5\textwidth]{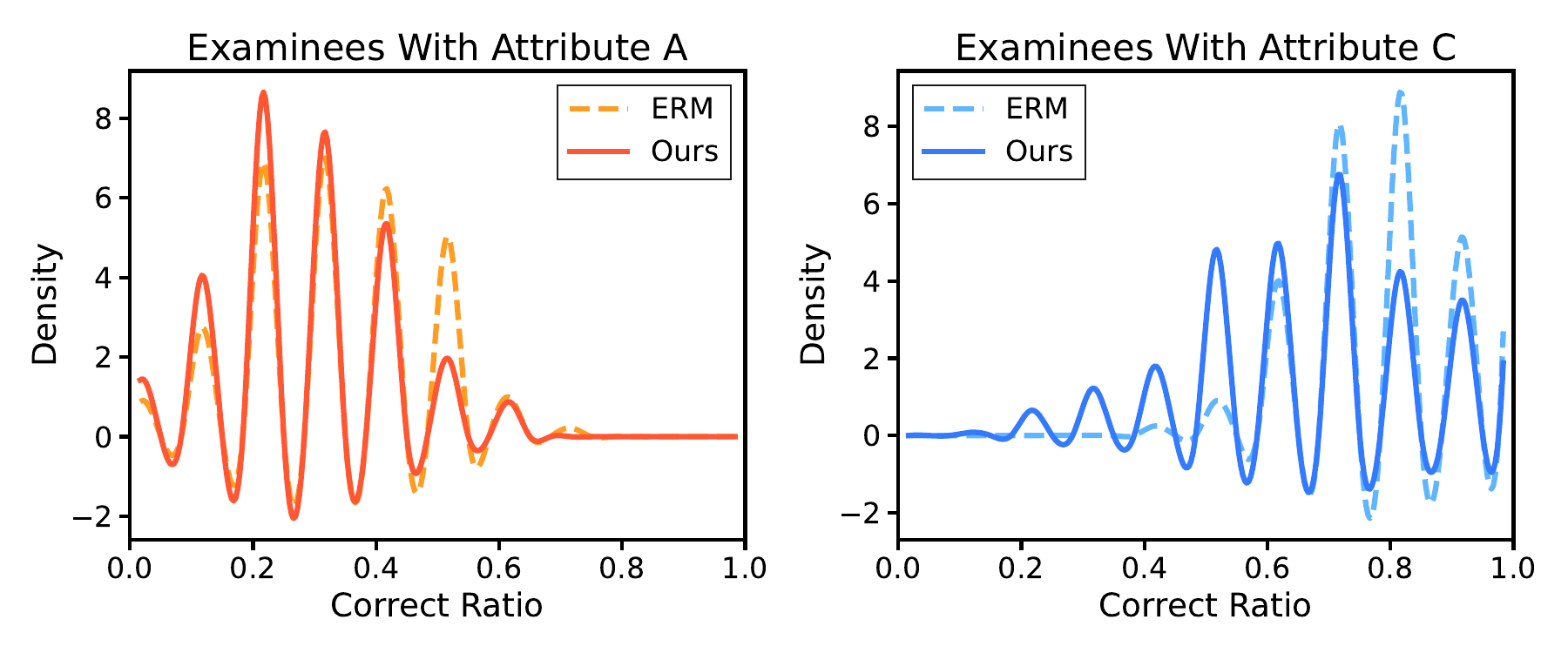} 
    \caption{Distribution of selected-question correct-response ratios across examinees.}
    \label{fig:sel_dis}
\end{figure}

While previous results demonstrate the effectiveness of our method in improving Question Selection Module performance under OOD conditions, we further investigate its mechanism by analyzing the distributional differences it introduces. Specifically, Figure~\ref{fig:sel_dis} presents the distribution of correct-response ratios among examinees under both the ERM baseline and our method. For examinees in attribute A, our method shifts the distribution toward a more moderate range (around 0.4), indicating that fewer overly difficult questions are selected and the risk of over-challenging is reduced. For examinees in attribute C, the skewness toward extremely easy questions is effectively alleviated, with the distribution centered around 0.6 instead of approaching 1.0 as in ERM, suggesting less overfitting to bias-aligned patterns overall.

Overall, these results demonstrate that our method effectively reduces the gap between the selected-question distribution and the meta set, yielding a more balanced and representative selection across examinees. By enriching bias-conflicting samples, it better preserves individual characteristics while aligning selection with a fairer target distribution, ensuring that proficiency estimation reflects true ability rather than artifacts of imbalanced training data. In turn, this leads to improved generalization, greater robustness, and more reliable assessment outcomes.  

\section{Conclusion}

In this study, we identify and address the problem of selection bias in data-driven Computerized Adaptive Testing, focusing on the imbalance of meta-set distributions during training. We propose a Mixup strategy guided by moderately performing examinees to augment underrepresented interactions and enhance the generalization of the Question Selection Module. The method is simple, effective, and can be seamlessly integrated into data-driven CAT training pipelines. Experiments on two benchmark datasets demonstrate that our approach mitigates bias while improving accuracy. By enriching bias-conflicting samples through synthetic data generation, the model learns more robust and less biased decision boundaries. This work highlights the importance of bias mitigation in intelligent educational systems and suggests addressing bias directly within the selection process as a promising direction for future research. Further investigations may explore integrating debiasing strategies with other CAT components to achieve broader improvements in reliability and validity.

\section{GenAI Usage Disclosure}
This paper used generative AI tools (e.g., ChatGPT) for non-substantive assistance such as language polishing, grammar correction, simple code generation, and writing suggestions. All technical content, experiments, and core contributions were conceived and implemented solely by the authors.

\begin{acks}
This research was partially supported by grants from the National Science and Technology Major Project (No.~2021ZD0111802), the National Natural Science Foundation of China (No.~62376086, 62406096), the Fundamental Research Funds for the Central Universities of China (No.~JZ2025HGTG0289), and the China Postdoctoral Science Foundation (No.~2024M760722). The computations were completed on the HPC Platform of Hefei University of Technology.
\end{acks}

\bibliographystyle{ACM-Reference-Format}
\balance
\bibliography{sample-base}
\end{document}